# Evidence for electronic signature of magnetic transition in topological magnet HoSbTe

**Authors:** Nana Shumiya[1], Jia-Xin Yin[1]†, Guoqing Chang[2]†, Meng Yang[3,4,5], Sougata Mardanya[6], Tay-Rong Chang,[6,7,8] Hsin Lin[9], Md Shafayat Hossain[1], Yu-Xiao Jiang[1], Tyler A. Cochran[1], Qi Zhang[1], Xian P. Yang[1], Youguo Shi[3,4,5], M. Zahid Hasan[1,10]†

**Affiliations:**

[1]Laboratory for Topological Quantum Matter and Advanced Spectroscopy (B7), Department of Physics, Princeton University, Princeton, New Jersey 08544, USA.

[2]Division of Physics and Applied Physics, School of Physical and Mathematical Sciences, Nanyang Technological University, Singapore 637371, Singapore.

[3]Beijing National Laboratory for Condensed Matter Physics and Institute of Physics, Chinese Academy of Sciences, Beijing 100190, China.

[4]Center of Materials Science and Optoelectronics Engineering, University of Chinese Academy of Sciences, Beijing 100049, China.

[5]School of Physical Sciences, University of Chinese Academy of Sciences, Beijing 100190, China.

[6]Department of Physics, National Cheng Kung University, 701 Tainan, Taiwan.

[7]Center for Quantum Frontiers of Research and Technology (QFort), Tainan, Taiwan.

[8]Physics Division, National Center for Theoretical Sciences, Taipei 10617, Taiwan.

[9]Institute of Physics, Academia Sinica, Taipei 11529, Taiwan.

[10]Lawrence Berkeley National Laboratory, Berkeley, California 94720, USA.

[11]Quantum Science Center, Oak Ridge, TN 37830, USA.

†Corresponding authors, E-mail: jiaxiny@princeton.edu; guoqing.chang@ntu.edu.sg; mzhasan@princeton.edu

**Topological insulators with intrinsic magnetic order are emerging as an exciting platform to realize fundamentally new excitations from topological quantum states of matter. To study these systems and their physics, people have proposed a variety of magnetic topological insulator systems, including HoSbTe, an antiferromagnetic weak topological insulator candidate. In this work, we use scanning tunneling microscopy to probe the electronic structure of HoSbTe with antiferromagnetic and ferromagnetic orders that are tuned by applying an external magnetic field. Although around the Fermi energy, we find minor differences between the quasi-particle interferences under the ferromagnetic and antiferromagnetic orders, deep inside the valance region, a new quasi-particle interference signal emerges with ferromagnetism. This observation is consistent with our first-principles calculations indicating the magnetism-driven transition of the electronic states in this spin-orbit coupled topological magnet.**

Recently, the field of quantum materials, especially strongly correlated and topological materials, has seen an explosion in research efforts and interest due to their exotic physics and potential applications [1-5]. Although researchers have been making tremendous progress to demonstrate a large number of topological systems in real life, it remains a challenge to realize topological materials with a strong interaction between electrons which are expected to host richer physics [6, 7]. In particular, topological systems that host intrinsic magnetic order have received greater attention in the community. In topological magnets, the time-reversal symmetry is broken, and hence they are expected to host various exotic quantum phenomena. In fact, a variety of topological magnets have been studied, and remarkable properties have been revealed [8-12], such as negative flat band magnetism in kagome magnet $Co_3Sn_2S_2$ [9] and Landau quantization in $TbMn_6Sn_6$ [11].

The WHM systems (W = {Zr, Hf or La}, H = {Si, Ge, Sn or Sb}, and M = {O, S, Se or Te}) are theoretically predicted to be stacking of two-dimensional topological insulators with the inclusion of spin-orbit interaction, and nodal lines semimetals when the spin-orbit coupling is neglected [10]. In the recent past, there have been a significant number of studies on WHM compounds, which reported promising results both theoretically and experimentally [13-30]. In particular, ZrSiS has been experimentally demonstrated to be topological nodal lines semimetals by various experiments [13-15]. ZrSiSe and ZrSiTe have been confirmed to host nodal line fermions by ARPES measurements [17], and the unconventional floating band type of surface state on ZrSiSe has been visualized by the quasi-particle interference (QPI) technique [18]. Recently, LnSbTe (Ln = lanthanide) [13, 21-30], a special group in the WHM systems, has been receiving greater attention. Certain lanthanide elements are known to induce intrinsic magnetic order of the system, and hence LnSbTe appears as a promising platform to study exotic electronic quantum states at the nexus of magnetism and topology.

HoSbTe, a compound that belongs to the group of LnSbTe is theoretically proposed to be a weak TI [28, 29]. HoSbTe crystalizes in the tetragonal P4/nmm (No. 129) group with lattice constant a = b = 4.23Å and c = 9.15Å. It consists of Te-Ho bilayers square lattices sandwiched by the layers of Sb square lattices as shown in FIG. 1(a). The crystal tends to cleave along the Te plane as shown by the yellow area in FIG. 1(a), due to the weak bonding between Te planes. It has an antiferromagnetic ground state with phase transition temperature $T_N$ = 4.5 K [28]. The magnetic field applied along the *c*-axis can drive a spin-flip transition to ferromagnetic order for a field larger than 2T [28]. FIG. 1(b) and (c) show the first-principles calculations of the electronic band structure in momentum space with antiferromagnetic order and ferromagnetic order in the presence of spin-orbit coupling, respectively. The electronic band structure calculations are performed within the framework of density functional theory [31, 32] using the projector augmented wave [33] pseudopotentials as implemented in the Vienna ab-initio simulation package [34, 35]. The exchange-correlation effects are considered within the generalized gradient approximation (GGA) proposed by Perdew, Burke, and Ernzerhof [36]. To incorporate the relativistic effect, the spin orbit coupling is included self-consistently in all our calculations. Along with that, the GGA+U scheme [37, 38] is used to treat the strong electronic correlation from Ho 4f states with effective Hubbard potential $U_{eff}$ = 7.0 eV. The plane-wave basis sets used for the bulk calculations are determined by the energy cut-off of 500 eV, while the Brillouin zone integrations are performed over a sampling of 12 × 12 × 5 Γ−centered k-mesh [39]. The electronic energy minimization is done up to $10^{-8}$ eV tolerance. We used the experimental lattice constant throughout all our calculations. We obtained the tight-binding model Hamiltonian from the atom-centered Wannier functions using the VASP2WANNIER90 interface and Wannier90 code suite [40]. The wannier functions are constructed based on Ho d and f orbitals and for both Sb and Te, s and p orbital. We then calculated the QPIs based on JDOS of the surface Green's function [41]. As seen in FIG. 1(b, c), topological nodal lines with sharp linear dispersion are both observed in antiferromagnetic (FIG. 1(b)) and ferromagnetic (FIG. 1(c)) structures around the X point. The main difference between the ferromagnetic bands and the antiferromagnetic bands is the spin-splitting due to the ferromagnetism. We observe that the spin splitting is very large for the states around the Γ point but almost negligible for the states around the X point. This is due to the different orbital contributions of the states.

Here we use Te-orbital as the representative (FIG. 1(d)), the states around Γ are from out-plane orbitals ($p_z$), while the states around X are from in-plane orbitals ($p_x$ and $p_y$). Considering the surface atoms are ferromagnetic for both antiferromagnetic and ferromagnetic states, it is challenging to resolve the difference from the nodal lines. From the bulk band, we can see obvious Zeeman splitting for the giant pockets around Γ. In contrast, the Zeeman splitting for states around X points is very small. The main difference between AFM and FM phase in band structure are the Zeeman splitting. Since STM is surface sensitive, we propose to detect the difference between the antiferromagnetic and ferromagnetic states by deleting the surface states related to the pockets around Γ, as shown in FIG. 1(e, f). The bulk resonance pocket at Γ "disappears" on the top atom layer. Yet, we find that, at 400meV below the Fermi level, the DFT calculations show obvious differences between the AFM and FM states for the states from Γ pockets [in the red box in FIG. 1(e, f)]: FM states establish much larger Zeeman splitting. Therefore, in our work, we focused on the QPI difference between AFM and FM states blow the Fermi level. And we indeed captured the difference.

High-quality single crystals of HoSbTe were grown by the Sb-flux method. High-purity Ho, Sb, and Te were mixed in a molar ratio of 1:20:1. They were placed in an alumina crucible in a quartz tube and sealed under a high vacuum atmosphere. The tube was then put into the furnace and heated to 1373 K, at which it was held for over 15h. After that and a dwell time of 10h, the crucible was slowly cooled to 1073 K at a rate of 2 K/h. Finally, centrifugation was performed to separate the crystals from the excess Sb. After the crystals are successfully grown, single crystals of HoSbTe up to 3mm x 3mm x 1mm were cleaved mechanically in situ at 77 K in ultra-high vacuum conditions, and then immediately inserted into the microscope head, already at $^4$He base temperature at 4.2 K. For cleaved crystals, we explore surface areas over 5 μm × 5 μm to search for atomic flat surfaces. To perform the measurements, commercial scanning tunneling microscopy Ir/Pt tip were used in this study. All the data shown in the main text were taken at 4.2K.

Our atomically resolved scanning tunneling microscopy topographic image reveals its square lattice structure (FIG. 2(a)). To gain deeper insight into its surface state, we perform systematic spectroscopic imaging on a large Te layer. By taking the Fourier transform of the differential conductance map (FIG. 2(b)), we obtain the QPI data. The QPI data shows a 3 lined crossing shape signal (FIG. 2(c)), which can be seen even more clearly in 4-fold symmetrized QPI data in FIG. 2(d). The 4-fold symmetry is consistent with surface states measurement by angle-resolved photoemission spectroscopy [29].

To further explore its band feature, we measured QPI at 0T (antiferromagnetic order) and 4T (ferromagnetic order) external, out-of-plane magnetic field at three different energy values, 0mV, -200mV, and -600mV, as shown in FIG. 3(a-c), respectively. The magnetic field was applied under zero-field cooling, after which we carefully approached the tip to locate the same atomic-scale area for tunneling spectroscopy. For the field-dependent QPI data, we first withdraw the tip away from the sample, and slowly ramped the field to 4T. Then we reapproached the tip to the sample. We found the exact same atomic area by searching based on impurities and vacancies positioning on the surface. Our result shows the three lined crossing-shaped QPI at the Fermi energy ($E_F$) (FIG. 3(a)). The QPI at the $E_F$ together with calculated QPI [FIG. 3(d)] and momentum resolved density of states at the surface (FIG. 3(g)) following the bulk dispersion shows the existence of nodal rings. However, even though the calculation also expects a large trivial pocket around Γ, which is much larger in bulk, the QPI result indicates almost no such state exists on the surface. This result indicates that the topological states are kept while trivial states are filtered out on the surface at $E_F$. At -200mV, QPI shows one lined crossing shape, and then become two thick-lined crossing shape at -600mV. Although there are little differences between 0T and 4T at -200mV and 0mV, we observe a band appearance at the middle corner indicated by the red box. In order to characterize the detail of the band property, we performed a calculation on its surface state (FIG. 3(g-i)) and corresponding QPI images (FIG. 3(d-f)) at 0, -200, and -600mV for ferromagnetic and antiferromagnetic order. The calculated QPI and surface states support the small dependence of magnetic order between ferromagnetic and antiferromagnetic at 0mV (FIG. 3(d, g)) and -200mV (FIG. 3(e, h)). As shown in the electronic band structure in FIG. 1(b,c), at and near the Fermi energy, the difference, Zeeman splitting, the nodal line due to magnetization, and the linear dispersion, are too small to be distinguished. Thus, at 0 mV or -200 meV, all the QPI patterns between antiferromagnetic and ferromagnetic are very similar. Our calculation of the surface states at Fermi energy provides a 4-fold symmetric diamond-like shape, which has also been supported by the previous ARPES measurement [28]. On the other hand, at the lower energy, at -600 mV, the calculation indicates significant field dependence between ferromagnetic and antiferromagnetic as marked in the red box (FIG. (f, i)). At -600 mV, there are clear differences of the surface states from the pockets around Γ between the ferromagnetic and antiferromagnetic phases. This scanning tunneling microscopy observation provides experimental evidence consistent with the magnetic transition of the ferromagnetic and antiferromagnetic phases. The full data set of the QPI data can be seen in Ref. [42]

Our experimental observation and theoretical analysis of the surface states of HoSbTe microscopically uncovers a transition of the electronic states between the ferromagnetic and antiferromagnetic phases. Future characterization of the relationship between the electronic state and magnetic order transition will be important for further exploring the band structure in momentum space as well as the topological nature of HoSbTe and the WHM family.

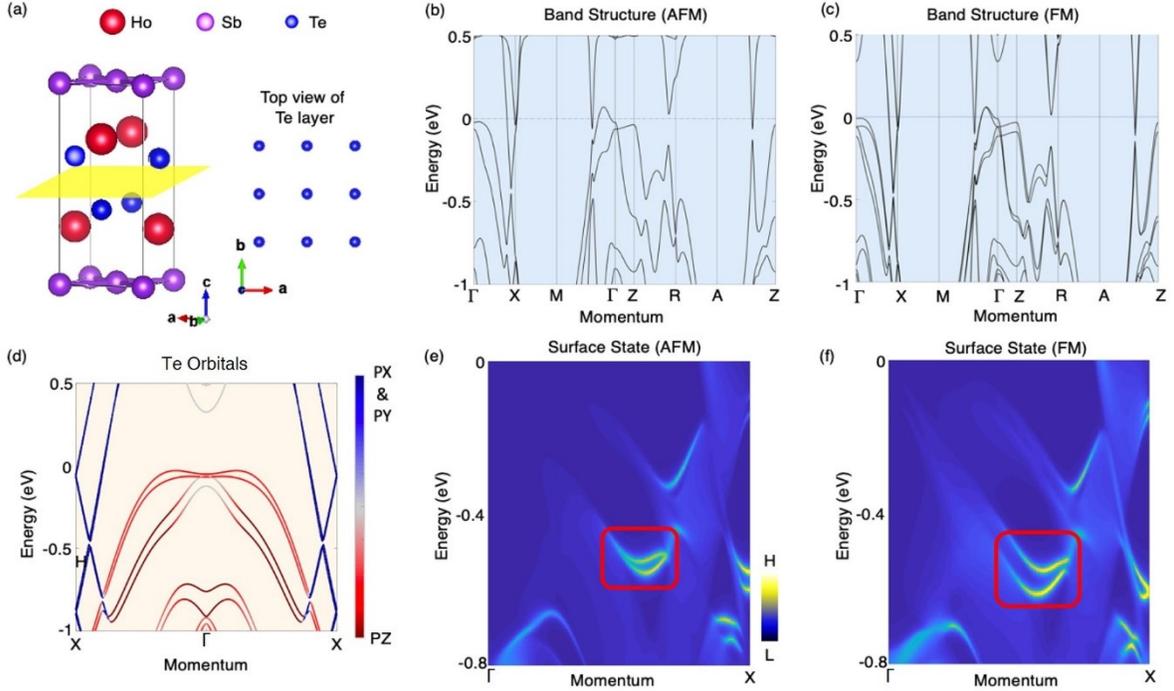

FIG. 1. (a) Crystal Structure of HoSbTe from three-dimensional view and top views. The yellow area denotes the natural cleaving termination. (b, c) DFT calculation of electric band structure of HoSbTe with antiferromagnetic order and ferromagnetic order, respectively. (d) Te orbital analysis. (e, f) Calculated surface state under antiferromagnetic order and ferromagnetic order, respectively. The difference was obtained by deleting the surface states related to the pockets around Γ. Red boxes indicates the states from Γ pockets.

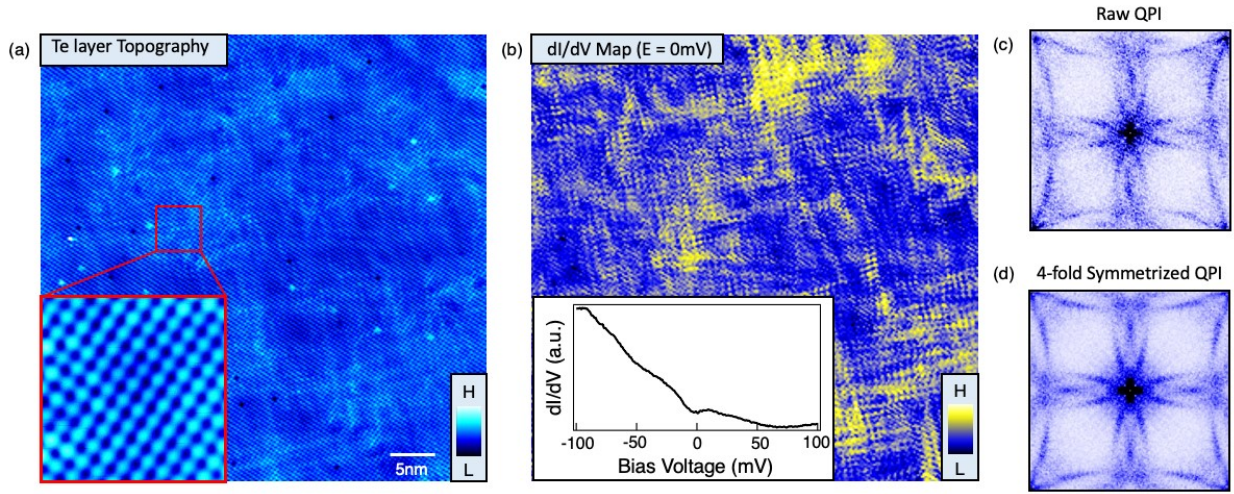

FIG. 2. (a) 50nm×50nm topographic image of Te layer surface. Inset: 5nm×5nm of atomically resolved topographic image of zoomed surface. (b) Corresponding dI/dV map taken under no external field at 0mV bias voltage at 4.2K. Inset: dI/dV spectrum on Te surface from -100mV to 100mV. The tunneling conductance data were obtained using standard lock-in amplifier techniques with a root mean square oscillation voltage of 0.3mV and a lock-in frequency of 973Hz. (c) Raw QPI taken from (b). (d) 4-fold symmetrized QPI taken from (b).

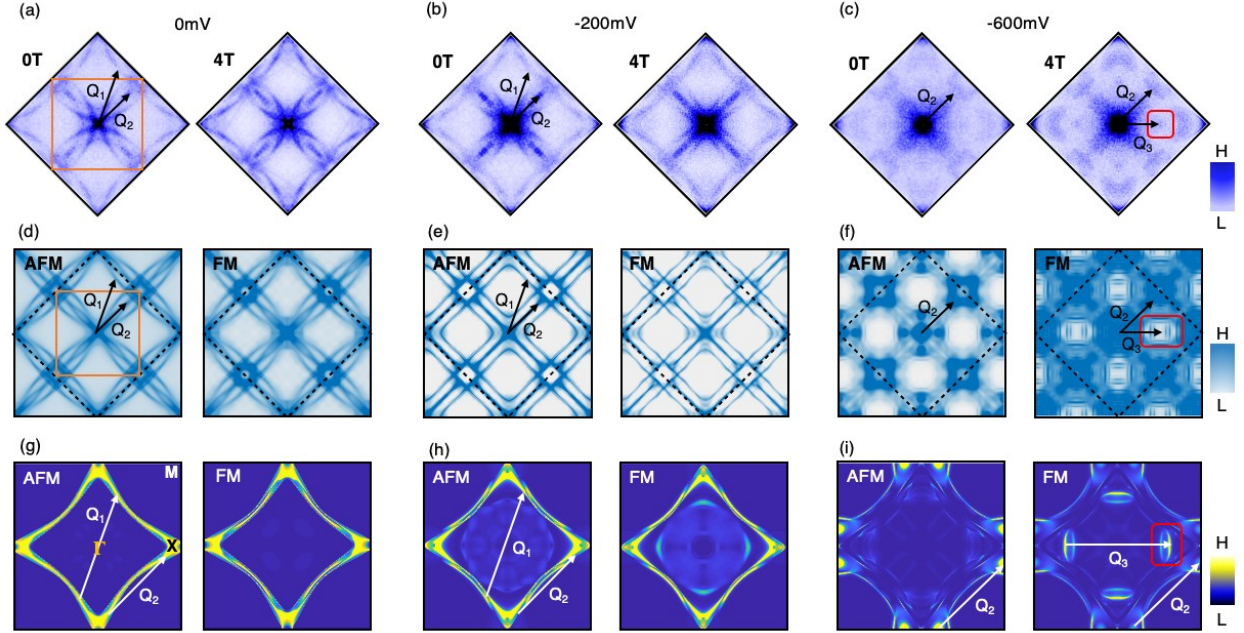

FIG. 3. (a-c) 4-fold symmetrized QPI at 0mV, -200mV, and -600mV, respectively. The orange square indicates the smallest Q space. Black arrows indicate Q vectors in the states. The red mark in (c) indicates the band that appeared uniquely with the application of the 4T external field. (d-f) Calculated QPI at 0mV, -200mV, and -600mV, respectively. Orange square, red marks, and black arrows correspond to those in (a-c). (g-i) Calculated surface states and $Q_1$ and $Q_2$ vectors corresponding to (d-f). White arrows indicate Q vectors in the states that correspond to the black arrows in (a-f), and the red mark indicates the surface band that corresponds to the surface states in (c) and (f).

**Acknowledgement:** M.Z.H. acknowledges support from the US Department of Energy, Office of Science, National Quantum Information Science Research Centers, Quantum Science Center and Princeton University. M.Z.H. acknowledges visiting scientist support at Berkeley Lab (Lawrence Berkeley National Laboratory) during the early phases of this work. Theoretical and STM works at Princeton University was supported by the Gordon and Betty Moore Foundation (GBMF9461; M.Z.H.). The theoretical work including ARPES were supported by the US DOE under the Basic Energy Sciences program (grant number DOE/BES DE-FG-02-05ER46200; M.Z.H.). G.C. acknowledges the support of the National Research Foundation, Singapore under its Fellowship Award (NRF-NRFF13-2021-0010) and the Nanyang Technological University start-up grant (NTUSUG). Y.S. acknowledges the National Natural Science Foundation of China (U2032204), and the K. C. Wong Education Foundation (GJTD-2018-01). T.-R.C. was supported by the Young Scholar Fellowship Program under a MOST grant for the Columbus Program, MOST111-2636-M-006-014, the Higher Education Sprout Project, Ministry of Education to the Headquarters of University Advancement at the National Cheng Kung University (NCKU), the National Center for Theoretical Sciences (Taiwan).

**Extended QPI measurements**

Extended QPI measurements were performed for field and temperature dependence using the same 50nm x 50nm clean area of Te surface. The field dependent data allows the comparison of electronic states between ferromagnetic order and antiferromagnetic order, and the temperature dependent data enables the comparison between antiferromagnetic and paramagnetic order. At 4.2K, we observed the emergence of corner states in QPI at high negative energies with field applied (FIG. S2) as comparison to QPI without field (FIG. S1), while for low energy and high positive energy, little field reaction has been observed.

At 4.7K, where the system hosts paramagnetic order, although the QPI at 0T (FIG. S3) and at 4T (FIG. S4) are expected to host the same surface states because of the same spin states, the data indicates a clear difference between QPI at 0T and QPI at 4T, especially at higher energies. This confliction suggests the system hosts antiferromagnetic spin fluctuations at the temperature (4.7K) that is only slightly above the transition temperature (4.5K). The difference in QPI vector shape still remain puzzling and is worth further study.

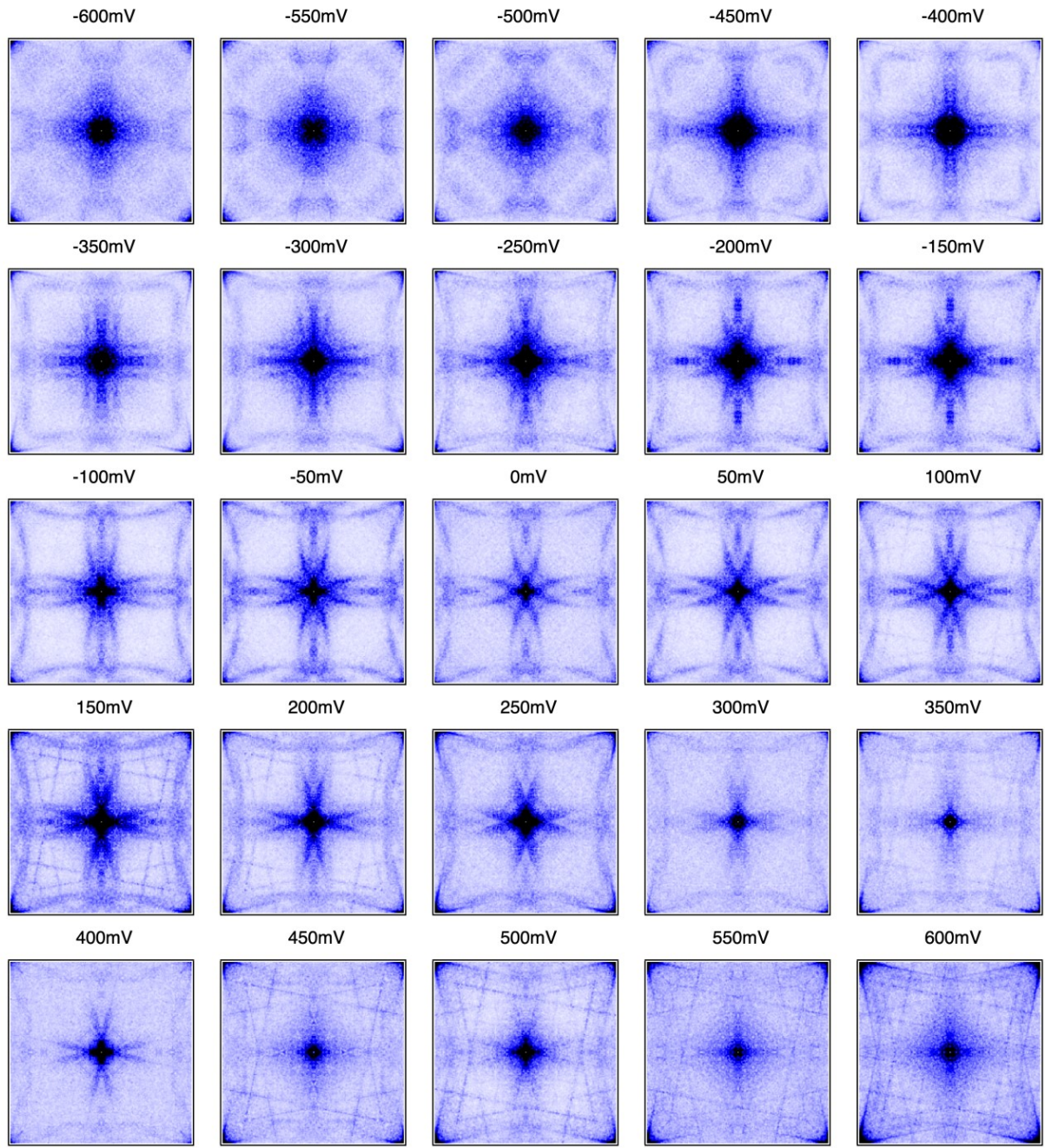

Supplementary FIG. 1. Symmetrized QPI of -600mV to 600mV under 0T at 4.2K

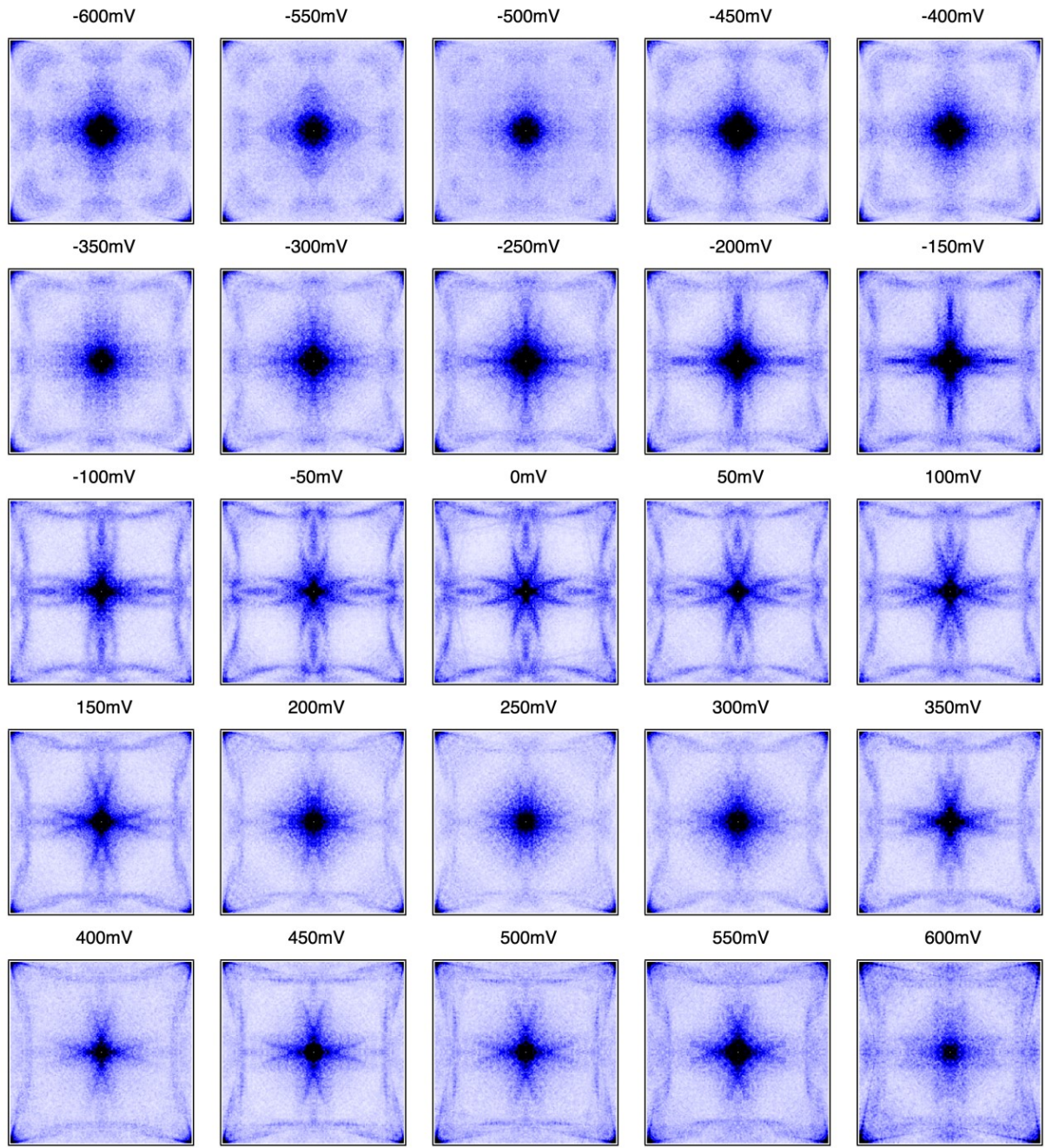

Supplementary FIG. 2. Symmetrized QPI of -600mV to 600mV under 4T at 4.2K

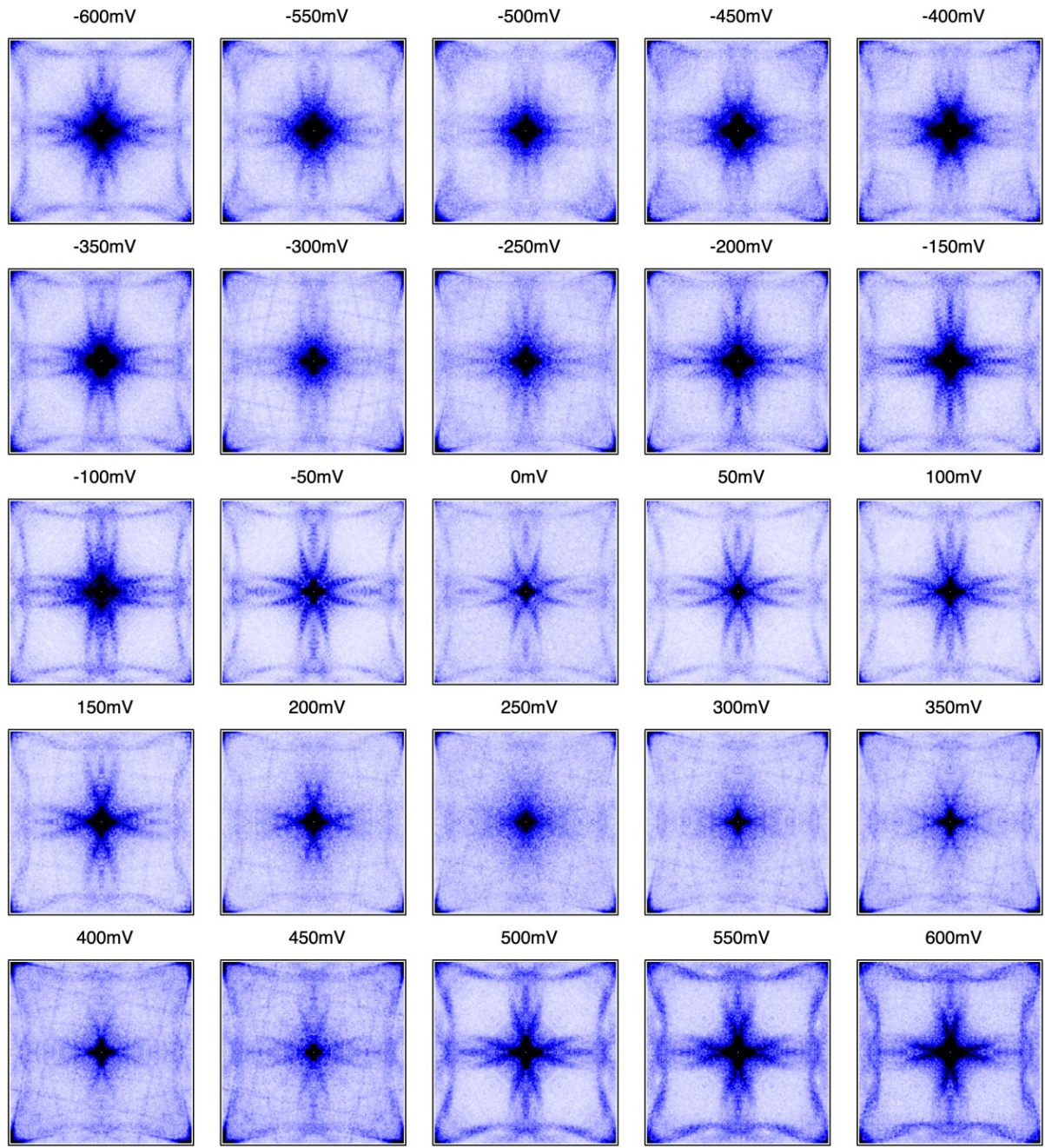

Supplementary FIG. 3. Symmetrized QPI of -600mV to 600mV under 0T at 4.7K

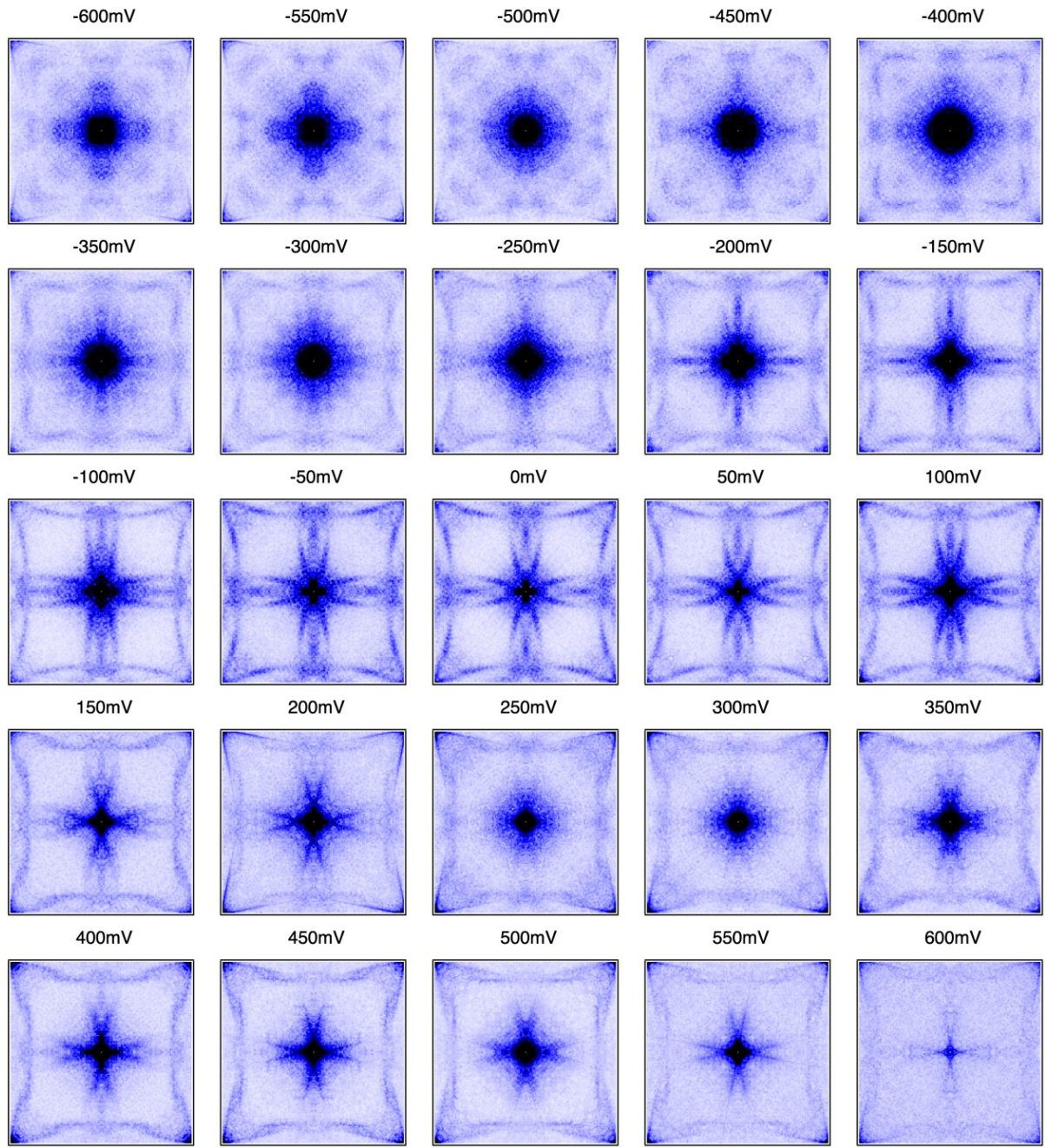

Supplementary FIG. 4. Symmetrized QPI of -600mV to 600mV under 4T at 4.7K